\documentclass{PoS}

\title{ The twisted Polyakov loop coupling and the search for an IR fixed point}

\ShortTitle{ The TPL coupling and the search for an IRFP}

\author{\speaker{Etsuko Itou}\\
        High Energy Accelerator Research Organization (KEK), Tsukuba 305-0801, Japan\\
        E-mail: \email{eitou@post.kek.jp}}

\abstract{We report the nonperturbative behavior of the twisted Polyakov loop (TPL) coupling constant for the SU($3$) gauge theories defined by the ratio of Polyakov loop correlators in finite volume with twisted boundary condition.
Carrying out the numerical simulations, we determine the nonperturbative running coupling constant in this renormalization scheme for $N_f=12$ SU($3$) gauge theory.

According to the perturbative two loop analysis, the $N_f=12$ SU($3$) gauge theory might have a conformal fixed point in the infrared region. 
However, the recent lattice studies show controversial results for the existence of the fixed point. 
We point out possible reasons of the controversy in previous works, and present our careful study. 
Finally, we find the infrared fixed point (IRFP) and discuss the robustness of the nontrivial IRFP of many flavor system under the change of the analysis method.
This talk is based on the paper~\cite{TPL}.}

\FullConference{31st International Symposium on Lattice Field Theory - LATTICE 2013\\
		July 29 - August 3, 2013\\
		Mainz, Germany}

\usepackage{amsmath}
\newcommand{\beq}{\begin{eqnarray}}
\newcommand{\eeq}{\end{eqnarray}}

\begin{document}

\section{Introduction}
Search for the infrared fixed point (IRFP) of SU($N$) gauge theories coupled to many fermions is one of the attractive applications beyond the QCD study~\cite{Lattice-plenary}.
The study on these theories who have the IRFP or nearly conformal properties has been motivated in the points of view of both the phenomenological model construction, {\it{e.g.}} a walking technicolor model or unparticle model,
and theoretical interest to reveal a novel universality class.
Recently there are many independent studies using the lattice simulation in the case of SU($3$) gauge theory coupled to $N_f=12$ fundamental fermions while the results for the existence of IRFP are controversy.
In this talk, I would like to consider the possible reasons for the controversy and to give a careful analyses to find IRFP.

Until now, four methods using the lattice numerical simulation have been applied to find the interactive IRFP for the gauge theories.
\begin{itemize}
\item Step scaling method

\item Hyperscaling for the mass deformed conformal gauge theory (mCGT)

\item Volume scaling for the Dirac eigenmodes

\item Shape of the correlation function of mesonic operator
\end{itemize}
The first one, called the step scaling method, is the method to measure the running coupling constants~\cite{Luscher:1991wu}.
The method is based on the renormalization group for the finite scaling, and it can be applied to the non-conformal field theories.
The second method uses the hyperscaling law for the mass deformed conformal gauge theory.
This method is based on the assumption of the existence of the interactive conformal field theory, where the scaling law of the mass spectrum of hadronic state or the chiral condensate in a small mass region can be fitted by  one universal value of the critical exponent, called mass anomalous dimension.
The application of the hyperscaling on the lattice was pointed out in the papers~\cite{Miransky:1998dh,Luty:2008vs, Sannino:2008pz}, and the detailed practical discussions of mass deformed conformal gauge theory have been shown in the paper~\cite{DelDebbio:2010ze}.
A similar method to find an IRFP and to estimate the mass anomalous dimension using the fit for the massless SU($3$) gauge theory has also been proposed~\cite{Patella:2012da, deForcrand:2012vh, Cheng:2013eu}.
The difference from the second method is that this third method can be applied to the massless fermion. 
The Dirac eigenmode for the massless fermion is the observable in this method. The scaling of ($1/L$), where $L$ is a finite lattice extent, tells us whether the theory is conformal or not in the infinite volume limit.  
The other independent method has also been suggested in the paper~\cite{Ishikawa:2013wf}. 
The authors gave a conjecture that the correlation function of the mesonic operator in the finite volume around the IRFP becomes the Yukawa-type function not an exponential form.

All methods have been applied to the SU($3$) $N_f=12$ theory.
However, the results for the existence of the IRFP are controversial even in the same method. 
The existence of the IRFP in $N_f=12$ theory was predicted by the perturbative beta function at $2$-loop~\cite{Caswell:1974gg} and higher~\cite{Vermaseren:1997fq} in the $\overline{\mathrm{MS}}$ scheme.
The phase structure of $N_f$ expansion was also studied in the paper~\cite{Banks:1981nn}.
Among the recent lattice studies, in Ref.~\cite{Appelquist:2009ty}, the running coupling constant was computed in the Schr\"{o}dinger functional
(SF) scheme~\cite{Luscher:1991wu,Luscher:1992an}, and exhibited scale independent behavior in the IR at coupling $g^{2 *}_{SF} \sim 5$.
And studies with the MCRG method~\cite{Hasenfratz:2010fi,Hasenfratz:2011xn}, studies on the phase structure in the finite temperature system~\cite{Deuzeman:2009mh,Miura:2011mc} and the scaling behavior with mass deformed theory~\cite{Appelquist:2011dp,DeGrand:2011cu,Aoki:2012eq} show the evidence of the IRFP. 
The volume scaling for the Dirac eigenmodes~\cite{Cheng:2013eu} and the shape of the mesonic correlation function~\cite{Ishikawa:2013wf, Ishikawa:2013tua} also show the signal of the IRFP.
In the studies of SU($3$) gauge theory with $N_f=7$ and $10$~\cite{Iwasaki:2003de, Hayakawa:2010yn,Appelquist:2012nz}, they found a signal of the conformality for each theory.
That  also suggests that $N_f=12$ theory is conformal.
On the other hand, the studies of the mass scaling behavior~\cite{Fodor:2011tu}
and the spectrum of the Dirac operator and the chiral symmetry~\cite{Fodor:2009wk, Jin:2012dw} show the evidence that this theory is not conformal at low energy.
This situation is confusing, since the existence of the fixed point is physical object and it is scheme independent quantity.

One possible reason for the controversial situation could be the underestimate of the discretization errors.
The nonperturbative running coupling constant using lattice simulations can be obtained using the step scaling method.
This method is established by the paper~\cite{Luscher:1991wu} and the nonperturbative running of  the renormalized coupling constant in the continuum limit can be obtained.
One important point which one should bear in mind is that the careful continuum extrapolation and estimation of the systematic uncertainty are important in the low $\beta$ ($\beta \equiv 6/g_0^2$ where $g_0$ is the bare coupling constant) region. 
However, there is no study of the running coupling constant which takes care of the discretization error carefully at least in the case of $N_f=12$.
For example, in the paper~\cite{Appelquist:2009ty}, the constant continuum extrapolation is taken.
That means the discretization effects, which is the renormalization scheme dependent, is neglected.

Another reason may be the bad choice of the value of $\beta$ and the mass of fermion.
In several previous works, the specific value of $\beta$ is chosen without any reasons.
In the lattice gauge theory with many flavor improved staggered fermion, it was reported that there is a new bulk phase in the strong coupling regime~\cite{Jin:2012dw, Deuzeman:2012ee, Cheng:2011ic}.
Furthermore, the existence of chiral broken phase in the strong coupling limit for the SU(3) theory with $N_f \le 52$ is also reported~\cite{deForcrand:2012vh}.
If the simulation is performed within the bulk phase, it gives an unphysical results because these bulk and chiral broken phases are not connected with the continuum limit with asymptotically free ultraviolet fixed point.
Our lattice gauge action is defined to reproduce the continuum theory in the high $\beta$ limit.
We have to avoid such phases, the global parameter search and the determination of the parameter region which is obviously connected to the high $\beta$ region is needed.

We report a study of the phase structure and the running coupling constant for SU($3$) gauge theories with $N_f=12$.
Firstly, we study the phase structure of these theories with both analytical and numerical methods, and then compute numerically the running coupling constant with the twisted Polyakov loop (TPL) scheme in the weak coupling (deconfinement) phase. 
The TPL coupling was proposed by de Divitiis {\it et al.}~\cite{deDivitiis:1993hj} for the SU($2$) case, and we extend it to the SU($3$) theory.
This renormalization scheme has no $O(a)$ discretization error, which is of great advantage when we take the continuum limit.
Another advantage of this scheme is the absence of zero mode contributions thanks to the twisted boundary condition~\cite{'tHooft:1979uj}.
This regulates the fermion determinant in the massless limit, which enables simulation with massless fermions.
In this work, we take the continuum limit carefully, and show the existence of the IRFP in the $N_f=12$ theory if we include the systematic uncertainty coming from the continuum extrapolation.

\section{Twisted Polyakov loop (TPL) scheme}
For the lattice gauge theory, there are several useful renormalization schemes for the gauge coupling constant, {\it e.g.} the Schr\"{o}dinger functional (SF) scheme~\cite{Luscher:1991wu}, the potential scheme~\cite{Campostrini:1995aa}, the Wilson loop scheme~\cite{Bilgici:2009kh}, the Wilson flow (Yang-Mills gradient flow) scheme~\cite{Fodor:2012td, Fritzsch:2013je} etc.
Twisted Polyakov loop (TPL) scheme is one of nonperturbative renormalized coupling schemes defined in finite volume.
This scheme is given in Ref.~\cite{deDivitiis:1993hj} in the case of SU(2) gauge theory, choosing the ratio of Polyakov loop expectation values for twisted and untwisted directions.
The advantage of this scheme is that there is no $O(a)$ discretization error and the leading contribution comes from $O(a^2)$.
We extend the definition in Ref.~\cite{deDivitiis:1993hj} to the SU(3) case. 
Although this scheme can be defined in the continuum finite volume, in this section we start a brief review of the definition of TPL scheme on the lattice.
\subsection{The definition of TPL scheme in the SU(3) gauge theory}
To define the TPL scheme, we introduce twisted boundary condition for the link variables ($U_\mu$) in $x$ and $y$ directions  and the ordinary periodic boundary condition in $z$ and $t$ directions on the lattice.
\beq
U_{\mu}(x+\hat{\nu}L/a)=\Omega_{\nu} U_{\mu}(x) \Omega^{\dag}_{\nu}, 
\eeq for $\mu=x,y,z,t$ and $\nu=x,y$.
Here, $\Omega_{\nu}$ ($\nu=x,y$) are the twist matrices which have the following properties:
$ \Omega_{\nu} \Omega_{\nu}^{\dag}=\mathbb{I}$,
$ (\Omega_{\nu})^3=\mathbb{I}$,
$ \mbox{Tr}[\Omega_{\nu}]=0$, and 
$\Omega_{\mu}\Omega_{\nu}=e^{i2\pi/3}\Omega_{\nu}\Omega_{\mu}$, for a given $\mu$ and $\nu$($\ne \mu$).

In the system coupled with fermions, we also have to define the twisted boundary conditions for fermions.
To preserve the translational invariance on the lattice, we introduce a ``smell" degrees of fermion $N_s$~\cite{Parisi:1984cy}, which can be realized by an integral multiple of the number of color symmetry $N_c$. 
We identify the fermion field as a $N_c \times N_s$ matrix ($\psi^a_\alpha$($x$)), where $a$ ($a=1,\cdots,N_c$) and $\alpha$ ($\alpha=1,\cdots,N_s$) denote the indices of the color and smell. 
We can then impose the twisted boundary condition for fermion fields as
\beq
\psi^a_{\alpha} (x+\hat{\nu}L/a)= e^{i \pi/3} \Omega_{\nu}^{ab} \psi^{b}_{\beta} (\Omega_{\nu})^\dag_{\beta \alpha}\label{eq:fermion-bc}
\eeq
for $\nu=x,y$ directions.
Here, the smell index can be considered as a part of  ``flavor'' index, so that the number of flavors should be a multiple of $N_s$, in our case $N_s$ should be the multiple of $N_c=3$.

The renormalized coupling in the TPL scheme is defined by taking 
a ratio of Polyakov loop correlators in the twisted ($P_x$) and untwisted ($P_z$) directions:
\beq
g^2_{\mathrm{TPL}}=\lim_{a \rightarrow 0} \frac{1}{k_{latt}} \frac{\langle \sum_{y,z} P_{x} (y,z,L/2a) P_{x} (0,0,0)^{\dag} \rangle}{ \langle \sum_{x,y} P_{z} (x,y,L/2a) P_{z} (0,0,0)^{\dag} \rangle }.\label{TPL-def}
\eeq 

At tree level, this ratio of Polyakov loops is proportional to the bare coupling.  
The factor on the lattice ($k_{latt}$) is obtained by analytically calculating the one-gluon-exchange diagram.
The Feynman rule for the SU($N_c$) gauge theory on the lattice with the twisted boundary condition is given in Appendix B in the paper~\cite{deDivitiis:1993hj}.
The value of $k_{latt}$ is given as
\beq
k_{latt}=\frac{1}{g^2 N_c}\frac{1}{\hat{L}^2} \sum_{\hat{k}_\mu} \frac{\exp(i \hat{k}^{ph}\cdot \hat{r})}{\sum_\mu \sin^2(\hat{k}_\mu /2)},
\eeq
where $\hat{L}=L/a$, $\hat{r}=(x,y,z,L/2a)$ and $\hat{k}_\mu$ denotes the momentum in each direction.
In the twisted direction, $\hat{k}_{\mu}$ is given by the sum of  the physical and the unphysical twisted momenta:
\beq
\hat{k}_{x,y} &=& \hat{k}_{x,y}^{ph}+\hat{k}_{x,y}^{\perp}=\frac{2\pi n_{x,y}^{ph}}{\hat{L}}+ \frac{\pi (2m_{x,y}^{\perp}+1)}{3\hat{L}}, \mspace{15mu}
\hat{k}_{z,t}  = \hat{k}_{z,t}^{ph}=\frac{2\pi n_{z,t}^{ph}}{\hat{L}}, \label{eq:k-mu}
\eeq
where $n_\mu^{ph}=0,\cdots \hat{L}/2-1$ and $m_{x,y}^{\perp}=0,1,\cdots, N_c-1$ with $(m_x^\perp,m_y^\perp) \ne (0,0)$.
The momentum $\hat{k}^\perp$ can be identified as the color degree of freedom $(N_c^2-1)$ in the color basis (see the Appendix B in the paper~\cite{deDivitiis:1993hj}).

\subsection{A fake fixed point of the TPL coupling constant}\label{sec:quenched-phase-str}
In this subsection, we would like to show a fake fixed point in the TPL scheme.
The fake fixed point is a kind of coordinate singularity in the theory space, and the existence of the fake fixed point depends on the renormalization scheme while the true fixed point of renormalization group is independent of the renormalization scheme.
The TPL coupling constant is defined by taking the ratio of the correlators of Polyakov loop in the twisted and the untwisted directions.
If the theory is in the confinement phase the correlation length of the Polyakov loop is shorter than the volume, and the gluon does not feel the boundary effect.
In such a situation, we can expect that the ratio of the Polyakov loop correlators becomes unity, and give a fake fixed point.
For this reason, it is awkward to extract the running coupling and try to give a physical meaning to it in such region.
The quenched QCD theory shows the confinement/deconfinement phase transition in the finite volumes, and we can use the TPL running coupling only in the deconfinement phase, where the magnitude of the Polyakov loop shows nonzero values.

To see the property of the TPL coupling in both confined and deconfined phases, we study $\beta$ dependence of the coupling constant at fixed lattice sizes. 
Apart from discretization errors, the coupling increases as $\beta$ decreases at a fixed lattice size.
In this test, we use smaller lattice sizes, $L/a = 2$ -- $6$, with 
relatively low $\beta$ values.
The configurations are generated by the hybrid Monte Carlo algorithm with the Wilson plaquette gauge action.
We measure the Polyakov loop and its correlator for every Monte Carlo trajectory, and each data has the same statistics of $20,000$ trajectories.

\begin{figure}[h]
\begin{center}
  \includegraphics*[height=7cm]{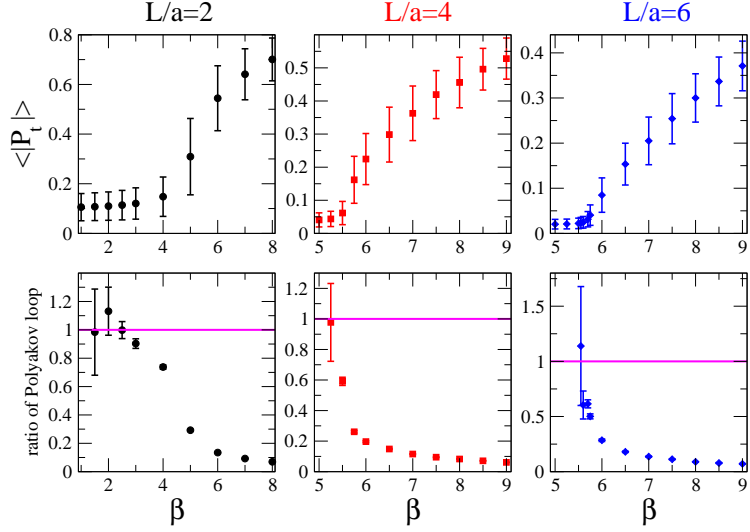} 
\caption{The ratio of Polyakov loop and the absolute value of Polyakov loop in $t$ direction for $L/a=2,4$ and $6$ .}
\label{fig:quenched-FLa}
\end{center}
\end{figure}

The TPL coupling constant and the absolute value of the Polyakov loop in $t$-direction are presented in Fig.~\ref{fig:quenched-FLa}.
The top panels denote the absolute values of the Polyakov loop and the bottom ones denote the corresponding TPL coupling scaled by the coefficient $k_{latt}$ for each lattice size.
We found that the absolute value of the Polyakov loop approaches zero in the low energy region.
The confinement/deconfinement phase transition occurs at the transition point of $\beta$ which depends on the lattice sizes.
From the bottom panels, we can see the ratio of Polyakov loop ($k_{latt} g^2_{\mathrm{TPL}}$) becomes unity below the transition point.

Since there can be a fake fixed point due to confinement, there is a question whether we can use this TPL scheme for the conformal fixed point search in IR region.
One way to judge that the fixed point is not the fake one is to check the the value of renormalized coupling.
Assume that a theory has IRFP. 
The fake fixed point appears at $g^2_{\mathrm{TPL}} \sim 1/k \sim32$.
If there is an IRFP at $g^{2*} _{\mathrm{TPL}} \ne1/k \sim 32$, then we can tell that the fixed point as a physical fixed point.
The other important check is to see the phase structure of the theory at the same time.
At the true conformal fixed point, the theory must be in the deconfinement phase.

\section{Phase structure of our simulation setup for $N_f=12$ SU($3$) gauge theory}
To avoid the bulk phase and search for the available region to use the TPL scheme,  we would like to reveal the phase structure of our lattice setup before studying the running coupling constant.
In the subsecntion \ref{sec:mass-phase} and \ref{sec:phase_ploop}, we observe the plaquette and the Polyakov loop in the broad range of $\beta$ -- $ma$ space, and we find that there is a bulk phase transition around $\beta < 4.0$ in the massless limit in our lattice setup.
In the subsection \ref{sec:Dirac-eigenvalue}, we show the eigenvalue of the Dirac operator for the massless fermions. According to the Banks-Casher relation, the chiral symmetry seems to be preserved in the weak coupling region.

\subsection{Plaquette values on the $\beta-ma$ plane}\label{sec:mass-phase}
\begin{figure}[h]
\begin{center}
   \includegraphics*[height=7cm]{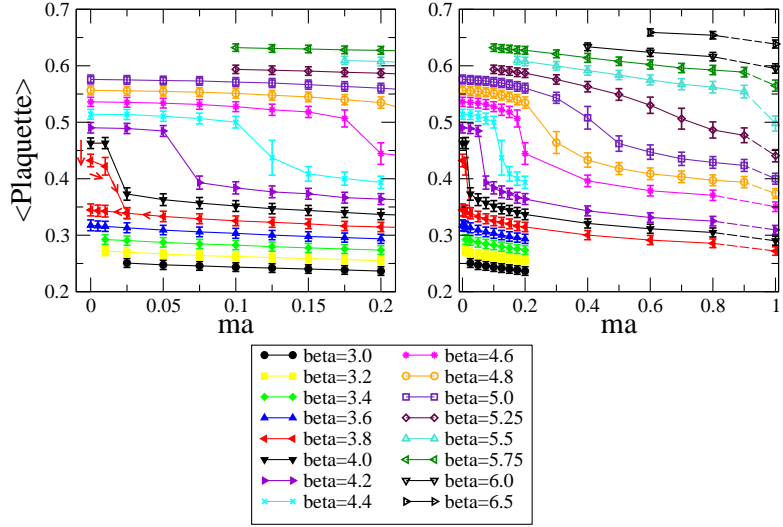}
  \end{center}
\caption{Plaquette values for several $\beta$ and $ma$ on $(L/a)^4=4^4$ lattice. The data at $ma=1$ on the right panel denotes the quenched QCD. The small (red) arrows near the massless at $\beta=3.8$ on the left panel shows the detailed history of the thermalization. The other data are thermalized from massive to massless direction.}
\label{fig:Plaq-L-4}
\end{figure}

Let us investigate the plaquette values of the $\beta$--$ma$ plane.
The left panel of figure~\ref{fig:Plaq-L-4}  shows the plaquette values on $(L/a)^4=4^4$ lattice in the $\beta$--$ma$ plane in the range of $0 \le ma \le 0.2$.
Most of the configurations are thermalized from massive to massless direction except for the small mass region in the $\beta=3.8$. 
The small (red) arrows near the massless at $\beta=3.8$ on the left panel in Fig.~\ref{fig:Plaq-L-4} shows the detailed histories of the thermalization.
We find that there are two different values in the $0 \le am \le 0.0125$ region.
The configurations giving larger values of the plaquette at the same $ma$ are generated starting from configuration with massless fermions in $\beta=4.0$; on the other hand those giving smaller values are obtained starting from the configuration with massive fermions at fixed $\beta$.
The hysterisis clearly indicates that there is a first order phase transition around this region.
At $\beta=4.0$ and $\beta=3.6$, there is no dependence on the thermalization process.

We also study larger mass region.
The right panel in Fig.~\ref{fig:Plaq-L-4} is the same plot as the left one for a broader region of $ma$.
In the quenched limit, we know that there is the first order phase transition.
In the figure, we plot the data for the quenched lattice at $ma=1.0$.
The gap seems milder in the larger mass region.

We also investigate the first order phase transition by changing the lattice volume.
There are slight differences between $L/a=4$ and $L/a=8$ for the critical value of $\beta$ in the massive region.
On the other hand, for the massless region, the lattice volume dependence is not clear at least the present interval of $\beta$ and $ma$ ($\Delta \beta=0.2,\Delta ma =0.01$ -- $0.025$)
Since the massless simulation at $\beta=3.8$ needs extremely finer molecular-dynamics time step size than $\Delta \tau =0.002$ ($\tau=1$ is $1$ trajectory), practically we could not generate the data.
The position of $\beta$ where the simulation becomes quite costly is the same for both $L/a=8$ and $L/a=12$.
It suggests that near the massless region there is a bulk phase transition in $\beta < 4.0$.

\subsection{Polyakov loop}  \label{sec:phase_ploop} 
Next, let us investigate the Polyakov loop.
Since the dynamical fermions breaks the center symmetry explicitly, there is no clear order parameter for the deconfinement phase transition. 
However, here we use the word ``deconfinement" or weak coupling phase for the region in the theory space where magnitude of Polyakov loop is clearly nonzero on the lattice.

According to the semi-classical analysis, we found that the true vacua is the one that the Polyakov loops in the untwisted directions have the nontrivial phase (See Sec.~4 in the paper~\cite{TPL}).
We also observe the real part of Polaykov loop in $t$-direction in our numerical simulation.
We can find that there is a gap of the real part of the Polyakov loop at fixed $ma$ data, and the value of $\beta$ at the gap corresponds to the critical value of $\beta$ of confinement/ deconfinement phase transition.
In the case of massless fermions on $L/a=4$ we find a gap at the $\beta=3.8$.
For $\beta$ smaller than the gap position the real part of the Polyakov loop is not consistent with zero, but it goes to zero continuously.
In the finite mass region, there is a weak jump, and the gap become larger in the smaller mass region.
The value of the critical $\beta$ in which the data shows the jump is the same with the plaquette study.

In the case of larger lattice volume ($L/a=8$), there is no clear jump in the case of $L/a=8$, but the real part of the Polyakov loop approach to the zero in the low $\beta$ region.
Again, the critical value of $\beta$ is the same with the plaquette study.
We can conclude that the phase for $\beta$ larger than the gap position can be identified as the deconfinement phase and that for $\beta$ smaller than the gap position is the confinement phase.

Finally, we study the phase structure of the massless fermion $N_f=12$ QCD for $\beta \ge 4.0$ and $L/a=6$--$20$.
We find that all configurations, which are used for  the running coupling constant study, live in the deconfinement phase (although it might be trivial since the transition seems to be the bulk and we concentrate on the parameter region within $\beta \ge 4.0$). 

\subsection{Dirac eigenvalue}\label{sec:Dirac-eigenvalue}
We also measure the eigenvalue of the Dirac operator for the massless configurations that we utilize to study for the running coupling constant (See Appendix F in the paper~\cite{TPL}).  
The data at the lowest $\beta$ for each lattice extent show the inconsistency with zero, and the $\beta$ dependence of the data at fixed lattice extent is smooth in whole $\beta$ region.
The scaling of the lowest eigenvalue for the lattice extent is proportional to $(1/L)$ not $(1/L^3)$, so that it is a signal of the chiral restoration even in the lowest $\beta$ in the weak coupling phase in our simulation.

\begin{figure}[h]
\begin{center}
  \includegraphics*[height=5cm]{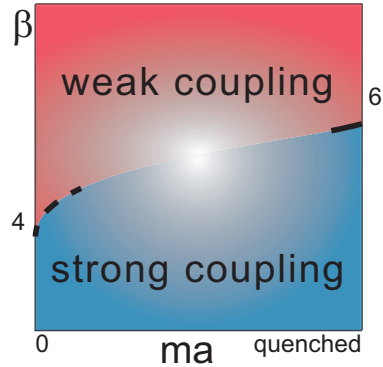} 
\caption{The phase structure of $N_f=12$ SU($3$) theory with naive staggered fermion. The dashed line denotes bulk phase transition and the solid line denotes the finite volume phase transition.}
\label{fig:phase-Nf12}
\end{center}
\end{figure}
The summary of the phase structure and the available region of the TPL coupling for the quenched and the massless $N_f=12$ QCD is the following.
Figure~\ref{fig:phase-Nf12} is a sketch of the phase structure for the naive staggered $N_f=12$ SU($3$) theory.
In the case of the quenched QCD, the correlation length becomes shorter in the lower $\beta$ region, and there is the finite volume phase transition where the theory goes to the confinement phase.
In the case of the massless $N_f=12$ SU($3$) theory, there is the similar behavior while the transition seems to be the bulk one at $\beta < 4.0$.
In the study on the running coupling constant in TPL scheme, we should focus on only the weak coupling phase on the lattice.

Furthermore, in $\beta \ge 4.0$ region with massless fermions, we also investigate the eigenvalue of Dirac operator.
The lowest eigenvalues are clearly nonzero even in the lowest $\beta$ for all lattice sizes, and the scaling of the eigenvalue for the lattice extent is proportional to $(1/L)$ not $(1/L^3)$.
It is a signal that the chiral symmetry is restored in $\beta \ge 4.0$.
We finally find the IRFP at higher $\beta$ values than the bulk phase transition point, although the values of physical critical $\beta$ at physical IRFP depend on the lattice sizes.
Our phase diagram (Fig.~\ref{fig:phase-Nf12}) is completely consistent with the conjectured phase diagram in the paper~\cite{deForcrand:2012vh} (Fig.~{10}).

\section{Step scaling function for $N_f=12$ SU($3$) gauge theory}
To investigate the evolution of the renormalized
running coupling, we use the step scaling method~\cite{Luscher:1991wu}.
Firstly we choose a value of the renormalized coupling $u{=}g^2_{\mathrm{TPL}}(\beta,a/L)$ at the energy scale $\mu=1/L$.
For each $L/a$ in the set of reference lattice size,
we find the value of $\beta$ which produces a given value of the renormalized coupling, $u$.
Then, we measure the step scaling function on the lattice
\beq
\Sigma (u,a/L;s){=} g^2_{\mathrm{TPL}}(\beta,a/sL)|_{g^2_{\mathrm{TPL}}(\beta,a/L)=u},
\eeq
at the tuned value of $\beta$ for each lattice size $sL/a$.
Here, $s$ is the step-scaling parameter.
The step-scaling function in the continuum limit $\sigma (s,u)$ 
is obtained by taking the continuum extrapolation of $\Sigma (u,a/L;s)$:
\beq
\sigma (s,u)= \lim_{a \rightarrow 0} \Sigma(u, a/L;s)|_{g^2_{\mathrm{TPL}}(\beta,a/L)=u}.
\label{eq:cont}
\eeq
This step scaling function ($\sigma(s,u)$) corresponds to the renormalized coupling at the scale $\mu=1/sL$.
The step scaling function $\sigma(s,u)$ can be defined independently for each input renormalized coupling ($u$), and the growth rate $\sigma(s,u)/u$ becomes unity when there is a zero in the beta function.

At first, we will discuss the global behavior of the growth rate from the perturbative to the IR region.
The nonperturbative running behavior shows the signal of the conformal fixed point in the IR region.
Then, we focus on the low energy region only and derive again the step scaling function by using the data only in the strong coupling region.
We discuss the stability of the IR fixed point by considering several systematic uncertainties and derive the universal quantity for the exponent of the beta function around the IRFP.
Finally, we obtain the critical exponent of the $\beta$ function around the IRFP. 

\subsection{global fit analysis}
\begin{figure}[tb]
  \begin{center}
   \includegraphics[height=5cm]{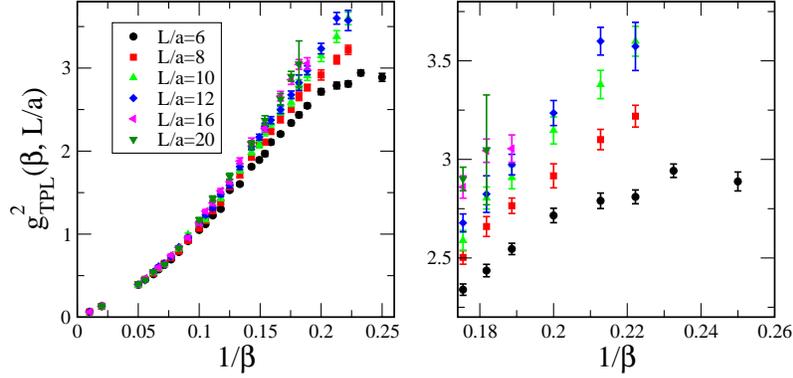}
  \end{center}
  \caption{TPL coupling for each $\beta$ and $L/a$. Left panel: Plots for the global region of $\beta$. Right panel: Plots only for the low $\beta$ region.} 
  \label{fig:global-g2}
\end{figure}

In Fig.\ref{fig:global-g2}, we show our simulation results for 
the renormalized coupling in TPL scheme as a function of $1/\beta$ 
for each $L/a$.
The left panel in the Fig.~\ref{fig:global-g2} shows a global behavior of the TPL coupling.
We can see the high energy behavior seems to be almost linear in $1/\beta$ as expected from the perturbation theory.
The right panel focuses on the low $\beta$ region.
In the low $\beta$ region for $L/a=6$ the TPL coupling has a maximum at $\beta=4.3$.
In contrast to the Schr\"{o}dinger functional scheme~\cite{Appelquist:2009ty}, the renormalized coupling gets larger for larger volume for all the range of $\beta$.
We consider that this difference comes from the lattice artifact which depends on the renormalization scheme.
To remove the effect, the careful continuum extrapolation is necessary.

In this study, we take $s=1.5$, and denote 
$\sigma(u)\equiv\sigma(s{=}1.5, u)$ in the rest of 
this paper for simplicity. 
The set of small lattices is taken to be 
$L/a=6, 8, 10,12$, therefore,  we need values of 
$g^2_{\mathrm{TPL}}$ for $L/a=9, 12, 15,18$ 
to take the continuum limit in Eq.~(\ref{eq:cont}).
For $L/a=9, 15$ and $18$,  we estimate values of 
$g^2_{\mathrm{TPL}}$ for a given $\beta$ by the  
linear interpolation in $(a/L)$ using the data on the 
lattices $L/a= \{8, 10\}$, $\{12,16\}$ and $\{16, 20\}$, respectively.
To estimate the systematic error of these interpolations, 
we also performed the linear interpolation in $(a/L)^2$, 
and found that the difference in the results with interpolations in $a/L$ and $(a/L)^2$ is negligible.

\begin{figure}[h]
\begin{center}
\includegraphics*[height=5cm]{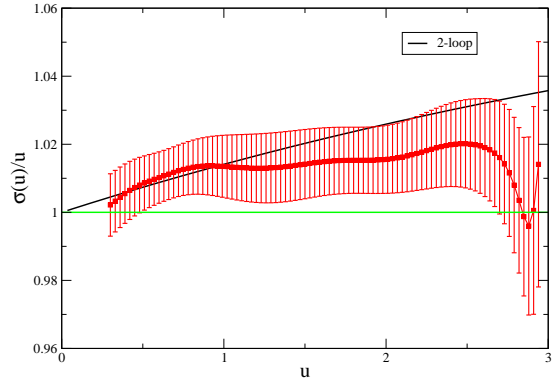}
\end{center}
\vspace{-0.5cm}
\caption{The growth rate $\sigma(u)/u$ as  a function of $u$ with statistical error. 
Two-loop perturbative value (black line) is also plotted for comparison. The horizontal (green) line denotes unity line, where the beta function is consistent with zero.} 
\label{fig:sigma-u-global}
\end{figure}
Now, we obtain the step scaling function explained above 
in a wide range of $u$. 
Figure~\ref{fig:sigma-u-global} shows the growth rate of the renormalized coupling ($\sigma(u)/u$) as a function of 
$u$ with statistical error which is estimated by jackknife method. 
We also carried out the bootstrap analysis independently, and found that the results are consistent with this jackknife analysis.

We found two things from this plot.
The first one is that the result is consistent with perturbation theory in the weak coupling regime. 
The TPL coupling coupling with this lattice set up looks promising under this analysis method.
The other one is the central value of $\sigma(u)/u$ becomes unity around $u = 2.7$, demonstrating the signal of a fixed point.
This is the first zero of the beta function from the asymptotically free regime.
It suggests the existence of an {\it infrared} fixed point around the region. 
Unfortunately, the upper values of the error bars do not cross the line $\sigma(u)/u=1$.
This means that we cannot exclude the possibility for the coupling constant to continue growing within the error bar.
We will investigate this quantity again by focusing only the strong coupling region and adding the data.
Furthermore, will give an estimation of the systematic error of the fixed point coupling in the next subsection.

\subsection{local fit analysis}
In the previous subsection, we found a signal of the IRFP around $u = 2.7$ from the global fit of the data.
Now we focus on the strong coupling region and will determine the fixed point coupling and the related universal quantity.
In this subsection, we take a narrow $\beta$ range in which $\beta$-dependence of $g^2_{\mathrm{TPL}}$ can be approximated by linear or quadratic functions of $\beta$.
We add more data to obtain the precise result and discuss the systematic uncertainties of the IRFP.

Practically, we will carry out the step scaling again with the data only in low $\beta$ region $u \ge 2.0$.
This region roughly corresponds to the range $\beta \le 7.0$.
The fitting function is chosen as a simple unconstraint polynomial function.
\begin{figure}[h]
\begin{center}
  \includegraphics*[height=5cm]{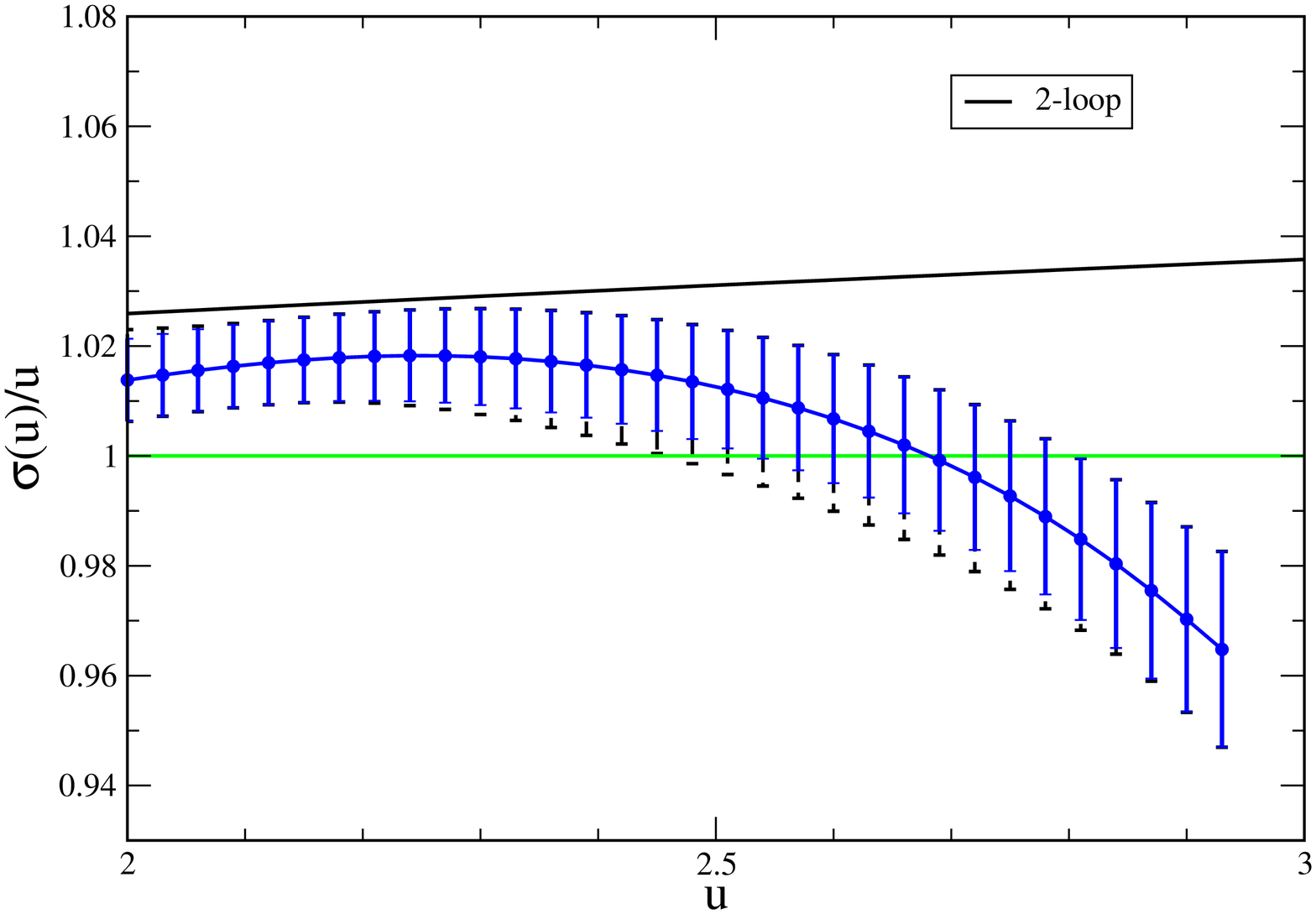} 
\caption{The local fit result for the growth rate of the TPL coupling. The solid (blue) error bar denotes the statistical error and the dot (black) error includes the systematic error. Two-loop perturbative value (black line) is also plotted for comparison. The horizontal (green) line denotes unity line, where the beta function is consistent with zero.}
\label{fig:sigma-u-local}
\end{center}
\end{figure}
We derived the step scaling function by using the same procedure as in the previous subsection.
The growth rate of the step scaling function is shown in Fig.~\ref{fig:sigma-u-local}.
As a central analysis with solid blue error bar, we take the four point linear extrapolation in $(a/L)^2$ with statistical error estimated by the jackknife method.
The dot (black) error bar includes the systematic error, which we will discuss later.
This local fit result clearly crosses the line $\sigma(u)/u=1$, which shows the existence of the IRFP.
Two central values in Figs.~\ref{fig:sigma-u-global} and \ref{fig:sigma-u-local} are consistent with each other within $1$-$\sigma$, despite the change of the data set, the fit range, and the fitting function.

Now, we would like to estimate the systematic error in our analysis.
\begin{figure}[h]
\begin{center}
  \includegraphics*[height=6cm]{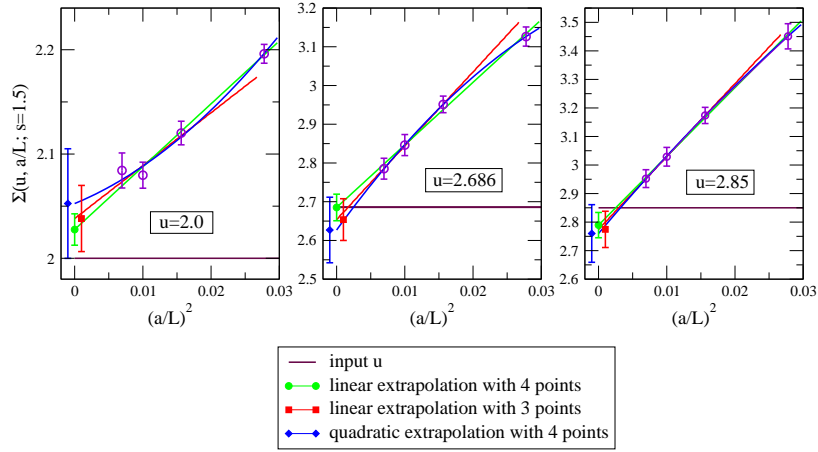} 
\end{center}
\caption{Continuum extrapolation for the case of 
input couplings $u=2.0$, $2.686$ and $2.85$. Each green line and the blue curve denotes the $4$ points linear  and quadratic extrapolation functions in $(a/L)^2$ respectively.
The red line shows the extrapolation function linear in $(a/L)^2$ for $3$ data points without the coarsest lattice data.
In the case of $u=2.0$, the step scaling function is larger than the input value, however, it becomes consistent with $u$ at $u=2.686$ and for the larger $u$ it is smaller than the input renormalized coupling constant.}
\label{fig:cont-lim-local}
\end{figure}
The dominant systematic error comes from the continuum extrapolation.
In Fig.~\ref{fig:cont-lim-local}, we show the comparisons of several types of continuum extrapolation for $u=2.0, 2.686$ and $2.85$.
As the central value, we take the linear extrapolation in $(a/L)^2$ for $L/a=6, 8, 10, 12$. 
We estimate the systematic error by taking the difference between the central value  and the result from linear extrapolation without the data on the coarsest lattice $L/a=6$.
Furthermore we compare the central value with the quadratic extrapolation with all the data at four values of $L/a$.
Figure~\ref{fig:cont-lim-local} shows the TPL renormalized coupling has 
a small systematic error in the strong coupling region, 
and all the values in the continuum limit agree within 1-$\sigma$ statistical errors.
The total error in Fig.~\ref{fig:sigma-u-local} is estimated by adding the difference between the continuum extrapolations as a systematic error to the statistical error in quadrature.
We conclude that the existence of the IRFP is stable in this analysis.

\subsection{Critical exponent}\label{sec:critical-exp}
Finally we obtain the critical exponent at the IRFP, which characterize the fixed point.  
In this theory, we have one irrelevant parameter, which is the renormalized coupling constant, around the nontrivial fixed point.
In the vicinity of the IRFP, the beta function for each renormalization scheme can be approximated by
\beq
\beta(u) \simeq - \gamma_g^\ast (u^\ast -u) +{\mathcal O} ((u^\ast-u)^2).
\eeq
Although the value of renormalized coupling at the IRFP is scheme dependent, we can easily find the coefficient $\gamma_g^\ast$ is the scheme independent quantity.

Now, we compute $\gamma_g^\ast$ from the slope of $\sigma(u)/u$ against $u$, and obtain 
$s^{-\gamma_g^\ast} = 0.79 \pm 0.11(\mbox{stat.})$ in the central analysis in the Fig.~\ref{fig:sigma-u-local}.  This leads to 
\beq
\gamma_g^\ast = 0.57^{+0.35}_{-0.31} (\mbox{stat.})^{+0} _{-0.16}\, (\mbox{syst.}),
\eeq 
where the first error is statistical error using the jackknife method and the second one is the systematic error from the continuum extrapolation estimated by the comparison to the $3$ point linear continuum extrapolation.
The value of $\gamma_g^\ast$ is sensitive to the variation of the slope, 
which causes rather large statistical error.
For the $s=2$ step scaling, the critical exponent of the beta function can be derived $\gamma_g^\ast=0.31^{+0.21}_{-0.18}(\mbox{stat.})$.
This is also consistent with our main results with $s=1.5$.

\section{Summary and discussion}
In the past several years, the study of many flavor SU($N$) gauge theories turned out to be attractive since whose infrared behavior is different from QCD.
In particular, more than $10$ independent groups have been studied the SU($3$) gauge theory coupled to $12$ flavor fermion within various methods while the results for the existence of the IRFP are controversial.
We consider that to give a final conclusion we should take care at least the following two points, {\it e.g.} the estimation of the continuum extrapolation and the careful understanding of the phase structure of each lattice setup.
We should avoid an unphysical bulk or chiral broken phases in strong coupling region if we search for the IRFP, since these phases might be not connected with the continuum limit with asymptotically free (ultraviolet) fixed point.
Futhermore, the critical point appears only at the continuum limit, so that we should estimate the systematic uncertainty carefully coming from the continuum extrapolation to give a conclusion of the existence of IRFP.
We have studied the phase structure by the observation of the plaquette, Polyakov loop and Dirac eigenvalues and then we have obtained the running coupling constant in the TPL scheme.
We finally found that there is a stable IRFP in our analysis.

\begin{figure}[h]
\vspace{1cm}
\begin{center}
  \includegraphics[width=6cm]{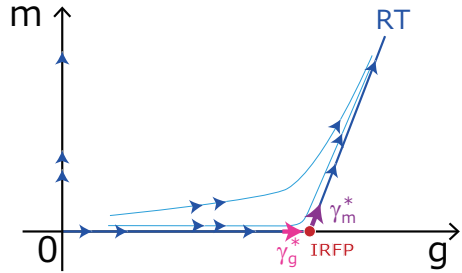}
 \caption{The theory space for the SU($3$) $N_f=12$ gauge theory.}
 \label{fig:theory-space}
\end{center}
\end{figure}

If there is no other relevant operator, then the renormalization group flows of the SU($3$) $N_f=12$ gauge theory are governed by the two dimensional theory spaces whose coordinates are the fermion mass and the gauge coupling constant (See: Fig.~\ref{fig:theory-space}).
The universal quantities to characterize the IRFP are the critical exponent of the beta function ($\gamma_g^*$) and the mass anomalous dimension ($\gamma_m^*$).

Our result for the critical exponent of the $\beta$ function is consistent with $2$-loop, 4loop ($\mathrm{\overline{MS}}$ scheme) within 1- $\sigma$ and also is consistent with 
the result in the SF scheme~\cite{Appelquist:2009ty} within 2- $\sigma$.
On the other hand, the mass anomalous dimension seems to be a discrepancy depends on the method.
The results which are obtained by the hyperscaling of the mCGT gave a larger value than the other results obtained by the volume-scaling and step scaling using massless simulation.
One possible reason of the discrepancy might come from the choice of value of $\beta$ and the mass of fermions in these simulations, and the lattice numerical data do not stay in the vicinity of the IRFP.
\begin{table}
\begin{center}
\begin{tabular}{|c|c|c|}
\hline
{} & $\gamma_g^*$ & $\gamma_m^*$\\
\hline
$2$ loop                      &  0.36          & 0.77\\
$4$ loop (MS bar)     &  0.28          & 0.25\\
Step scaling (SF scheme) Ref.~\cite{Appelquist:2009ty} &  0.13(3)      &          \\
hyperscaling I (mCGT) Ref.~\cite{Appelquist:2011dp} &  {} & 0.403(13) \\
hyperscaling II (mCGT) Ref.~\cite{DeGrand:2011cu} &  {} & 0.35(23)     \\
hyperscaling III (mCGT) Ref.~\cite{Aoki:2012eq} &  {} & 0.4 -- 0.5       \\
hyperscaling IV (Dirac eigenmode) Ref.~\cite{Cheng:2013eu} &  {} & 0.32(3)\\
Step scaling (our result) Ref.~\cite{TPL},~\cite{anomalous-dim} &  $0.57(35)$ & $0.044^{+0.062}_{-0.040} $\\
\hline
\end{tabular}
\caption{Current results for the critical exponents around the IRFP of $N_f=12$ SU($3$) gauge theory.
Note that in the papers~\cite{DeGrand:2011cu, Aoki:2012eq}  there is no `` $^*$ " on the gamma in their own papers. The value of the $\gamma_m^*$ in the hyperscaling IV is updated to $\gamma_m^*=0.25$ in Ref.~\cite{Hasenfratz:2013eka} }
\end{center}
\end{table}

\section*{Acknowledgements}
We would like to thank all ex-collaborators, in particular H.~Matsufuru, T.~Onogi and T.~Yamazaki for useful discussions.
Numerical simulation was carried out on
NEC SX-8 and Hitachi SR16000 at YITP, Kyoto University,
NEC SX-8R at RCNP, Osaka University,
and Hitachi SR11000, SR16000 and IBM System Blue Gene Solution at KEK 
under its Large-Scale Simulation Program
(No.~09/10-22, 10-16, (T)11-12, 12-16 and 12/13-16), as well as on the GPU cluster at Osaka University and 
Taiwanese National Centre for High-performance Computing.
We acknowledge Japan Lattice Data Grid for data
transfer and storage.
E.I. is supported in part by
Strategic Programs for Innovative Research (SPIRE) Field 5.
This work is supported in part by the Grant-in-Aid of the Ministry of Education (No. 
22740173). 


\end{document}